\documentclass[11pt,twoside]{article}

\usepackage{asp2004}
\usepackage{lscape}

\begin{document}
\title{Turbulent Elasticity of the Solar Convective Zone and the Taylor Number Puzzle}   
\author{Peter Todd Williams}   
\affil{\small{405 14${}^{\it th}$ Street, Suite 1207, Oakland, CA 94612 \\
Guest Scientist, Los Alamos National Laboratory}}    

\begin{abstract} 
Previous work on the angular momentum balance and meridional circulation of the solar convective zone (SCZ)
generally consists either of semi-analytic approaches in which a simple turbulence model is adopted, or full direct
numerical simulation (DNS) of the hydrodynamics or magnetohydrodynamics (MHD). In both
 instances the inclusion of magnetic fields has been troublesome. Also, both
approaches have had difficulty reproducing the known angular velocity profile of
the SCZ; this is the Taylor Number Puzzle.

I discuss preliminary work in which I incorporate magnetic fields into a
viscoelastic turbulence model for the SCZ and apply this to the problem of meridional
circulation and angular momentum balance. I suggest that such an approach may
help solve the Taylor Number Puzzle of the SCZ and bring theoretical predictions
for the large-scale motion of the SCZ in line with observations.
\end{abstract}



\thispagestyle{empty}
The Taylor number puzzle is this: in mean-field simulations of the combined system
of meridional circulation and rotation
of the solar convective zone (SCZ), one may come reasonably close to reproducing the observed
rotation profile by making the Taylor number artificially small. If on the other hand
one uses realistic values for the Taylor number, the meridional circulation is much larger in
magnitude than otherwise. This results in a theoretical angular momentum profile
that differs markedly from the observed profile. \citep{Braetal1990}

To be fair, the puzzle may have already been solved, at least to a degree.
That is because the situation changes if one allows for various
mechanisms that create a significant baroclinic vector, as baroclinicity appears as a source term in the meridional circulation.
This baroclinicity could be generated by latitudinal variations in the thermal transport, anisotropy of the thermal diffusivity
tensor, subadiabaticity of the solar tachocline, or some combination of these effects \citep{KicRud1995a, KicRud1995b, Rem2005, Mieetal2005}.
Another possibility is that additional Reynolds stresses due to the anisotropic kinetic alpha (AKA) effect may be important
to getting theory to match observation \citep{RekRud1998}. 
Still, I take an alternative point of view that the solution lies in part neither in baroclinicity nor in Reynolds stresses, but
in the physics of Maxwell stresses. Here I focus particularly on {\em turbulent} Maxwell stresses rather than integral-scale fields.

The provenance of the Taylor number is laboratory fluid dynamics.
The definition varies, but the most common expression given is that
\vskip - 2 mm
\begin{equation}
\mathrm{Ta} = \frac{4 \Omega^2 R^4}{\nu^2}.
\end{equation}  \pagebreak
${\mathrm{Ta}}$ is a dimensionless quantity used to characterize the relative
strength of centrifugal driving and viscous dissipation of meridional circulation, particularly in
cylindrical and spherical Couette flow. 
This latter problem, the flow between
two rotating spheres, bears some obvious similarities to the SCZ.
In the laboratory, a meridional circulation is driven when the spheres are differentially rotating.
The magnitude of this secondary flow relative to the primary azimuthal flow increases as 
$\mathrm{Ta}$ (or equivalently $\mathrm{Re}$) increases.

To understand what drives
the circulation in both cases, let us look at the vorticity equation for a compressible fluid
\begin{equation}
\partial_t\omega \simeq \nabla \times \left( u \times \omega \right) +  \frac{1}{\rho^2} \nabla\rho \times \nabla P +
\nu \nabla^2 \omega 
\label{eq:vort1}
\end{equation}
Actually, that is a bit of a lie; the viscous term for a compressible fluid is more troublesome
than indicated here, which is why I feel compelled to use a $\simeq$ instead of
$=$ above. Let us not worry about these details for the moment. 
The viscous term damps meridional circulation\footnote{Strictly speaking, this depends upon how
broadly one defines viscosity; here I am using the term in a narrow sense.}. 
In the case of the laboratory flow one could argue that viscosity also drives circulation through viscous coupling to
the spheres, creating an adverse angular momentum profile; that is not our concern here.
In the SCZ there are also contributions to the Reynolds stresses that are often referred to as 
non-diffusive viscous terms (this terminology reflects a broader sense of the word ``viscosity'' than
I am adopting here).
These stress components are very important to
the maintenance of differential rotation in the SCZ; this is the lambda-effect \citep{Rud1989}. The primary stresses due to the lambda-effect
are $r\phi$ and $\theta\phi$ stresses, which do not appear in the $\phi$ component of the vorticity equation. Rather, the influence
of the lambda-effect on meridional circulation is indirect. The lambda-effect bears on the meridional
transport of angular momentum, which contributes to the nonuniform angular velocity profile, which in turn generates
a meridional circulation through the $\nabla\times\left(u\times\omega\right)$ term. This is clearly
important for the SCZ, but again, it is not our concern here.

The physical nature of the advection/stretching term, the first term on the right-hand side (RHS) of eq.~({\ref{eq:vort1}}),
 is best seen in cylindrical coordinates. Then, in an inertial frame of reference,
when the meridional
circulation is negligible,
\begin{equation}
\nabla \times \left( u \times \omega \right)  = R \partial_z(\Omega^2),
\label{eq:cent}
\end{equation}
and this in turn is simply the curl of the specific centrifugal force in the locally corotating reference frame.
This tells us that there is a source of meridional circulation if
the rotation is not constant on cylinders.

For the measured rotation of the SCZ,
$\partial_z \Omega({R,z}) \,\not{\!\!\!\ll}\, \partial_R \Omega(R,z)$; the RHS of eq.~({\ref{eq:cent}}) is dynamically significant.
Under such circumstances, a barotropic fluid cannot undergo pure rotation and be in simultaneous gravitational, centrifugal, and pressure equilibrium. There must be a meridional circulation.
The sense of action of the centrifugal driving can be read directly off of graphs of isostrophic contours.
By inspection of the results of helioseismology, e.g. \cite{Schetal1998}, the predominant sign of $\partial_z(\Omega^2)$ in the northern hemisphere is negative. 
The centrifugal force should drive a counterclockwise (CCW) circulation in the usual notation (that is, towards the poles
at the surface and towards the equator at the base of the SCZ). This is similar to the driving of a secondary flow in spherical Couette flow when
the inner sphere is rotating faster than the outer sphere; at moderate Re this flow is dominated by a single large axisymmetric cell in each hemisphere,
CCW in the northern hemisphere.

Applied to the SCZ using the turbulent viscosity for $\nu$, the numerical value of $\mathrm{Ta}$ is very high;
\cite{RekRud1998} suggest $\mathrm{Ta} \simeq 10^{6\pm1}$. Turbulent viscosity is simply not effective at dampening
meridional circulation.
Absent additional physics in the SCZ, the angular velocity
profile as measured should strongly drive a meridional circulation. This circulation in turn should redistribute angular momentum, again through
the $\nabla \times (u \times \omega)$ term, which in the SCZ should drive the system to a Taylor-Proudman state. This is not seen.

We come now to the second term on the RHS in eq.~({\ref{eq:vort1}}). This term is of course irrelevant for the laboratory
case of spherical Couette flow, but probably important for the SCZ.
In principle, the baroclinic vector could balance out the centrifugal term in the meridional vorticity equation, creating a 
meridional flow much slower in magnitude than one would otherwise expect.
Of course, one is not free to adopt
whatever density and pressure profiles one wishes. In addition to the sought mechanical equilibrium, the system must also
satisfy thermal equilibrium. 
Sadly, the direction and magnitude of energy flux is more difficult to ascertain in the convective zone than it is in
the radiative zone. 
Simply note here that
for the baroclinic vector to have the right sign to balance the centrifugal forces, 
following an isopycnic surface, pressure and hence entropy must increase
as one moves from the equator to the poles.
The temperature gradient
must be less inclined to the axis of rotation than either the pressure or density gradients.  One might expect this to cause an 
observable latitudinal temperature difference. 
Various means by which Nature may create such baroclinicity and thus solve the puzzle 
were briefly referenced at the start of this paper.

As a hint for what is to come, note that the physics of turbulence is not included in eq.~({\ref{eq:vort1}}),
or at least not included very {\em well}.
Granted, it is common to think of turbulence as creating an effective viscosity, a notion that I have already
appealed to in writing the Taylor number for the SCZ in the first place. It is better, though, to write
the effect of turbulence in  eq.~({\ref{eq:vort1}}) applied to the SCZ
as the divergence of the turbulent stress tensor, which includes Reynolds stresses. \cite{Kitetal1994}
suggest an effect in convective turbulence in rotating, stratified fluids; this AKA-effect was mentioned previously.
It creates $r\theta$ and $\theta\theta$ Reynolds stresses, and so is a source term for meridional circulation.
\cite{RekRud1998} show that this effect can help solve the Taylor number puzzle as well.

That is where things stand in the literature, in an extremely small nutshell. What physics might be missing, if any? 
Certainly one piece of missing physics is that the sun can and does support magnetic fields.
Magnetic fields in turn create a stress, the Maxwell stress, the divergence of which is a force that acts upon the
fluid. None of this is taken into account in eq.~({\ref{eq:vort1}}). Naturally, it has long been recognized  that 
magnetic fields might be dynamically significant to the solar interior. The problem is that magnetic fields are difficult to treat.
Here I am going to discuss some recent work on a way of treating one particular component of the magnetic field --- the anisotropic, small-scale,
turbulent magnetic field --- and I am going to explain what I think is a very interesting reason to suspect that this might help
get the angular velocity profile of the SCZ right.

Let me take what appears to be a detour by returning to spherical Couette flow. Even an incompressible
fluid in spherical Couette flow, for the case we have considered where the inner sphere rotates
faster than the outer, can produce a reversed, clockwise (CW) flow rather than a CCW flow, {\em without the
help of any baroclinicity.}
All that is required is a viscoelastic fluid rather than a viscous one \citep{YamFujMat1997, YamMat1997}. 
In such a fluid, there is
an additional source of meridional vorticity that is not captured in eq.~\ref{eq:vort1}.

A viscoelastic fluid can be thought of as a viscous fluid where the stress has a decidedly finite (non-zero)
relaxation time $s$. On short timescales, the fluid behaves elastically; on long timescales the fluid behaves viscously.
Such behavior is
typically a result of the stress being due to long polymers rather than colliding point-like particles. For example,
bread dough made from wheat is viscoelastic because of the gluten polymer; likewise, the ovalbumin protein (a polymer)
makes egg whites viscoelastic.
In the case of steady shear, a viscoelastic fluid produces a streamwise elastic stress, creating
a positive (first) normal stress difference.
\footnote{The first normal
stress difference $N_1$ is the difference between the streamwise normal stress and the cross-stream
normal stress and is zero for a Newtonian fluid.}

The primary (azimuthal) flow in spherical Couette flow is a type of
of circular viscometric
\footnote{A viscometric flow is
a simple shear flow so that the relative-right Cauchy-Green tensor can be written as the
sum of the unit tensor and the first and second Rivlin-Eriksen tensors. An axisymmetric
flow with negligible meridional circulation is one example.}
flow. Almost universally, viscoelastic fluids
show this remarkable property in rotating viscometric flows: when viscoelastic effects dominate, 
there is a general tendency for the secondary (meridional) flows in such fluids to be
{\em opposite} to the secondary flows that would have been obtained for the centrifugally-driven
circulation of a purely viscous fluid.
This is known as secondary flow reversal.
It should be clear that a similar effect in the SCZ would slow the inertially-driven
meridional circulation, which is what we need.

For this picture of secondary flow reversal 
to be applicable to the SCZ, we require the dynamics of turbulence in the SCZ to be similar
to the dynamics of a viscoelastic fluid. But that is no problem. 
Both kinematically and dynamically, a turbulent, tangled magnetic field bears more semblance to a network of
polymers than it does to colliding point-like particles \citep{Ogil:2001, Will:2001, LonMcLFis2003, Ogil:2003, Will:2004}.
Let overbars denote averages and primes denote fluctuations. Neglecting the technical issues of Reynolds versus Favre averaging,
the turbulent stress $W_{ij}$ can be split into a Reynolds stress $R_{ij}$ and a turbulent Maxwell stress ${\cal{M}}_{ij}$,
this latter being the sum ${\cal{M}}_{ij} = M_{ij} - \frac{1}{2}M_{kk}\delta_{ij}$ where $4 \pi M_{ij} = \overline{B'_i B'_j}$.
 Let us suppose that the Reynolds stress will
consist of a turbulent viscosity and lambda-effect terms, as in previous mean-field work, and suppose as well
that the turbulent magnetic field will contribute visco{\em elastic} behavior, like a network of polymers. In principle,
pressure terms from both types of stress may contribute as well, but I am ignoring isotropic turbulent stresses here.

It is the normal stress difference $N_1$ (due to the elastic streamwise tension stress)
of a viscoelastic fluid under steady shear that is largely responsible for the secondary flow reversal in
viscometric flows such as our laboratory example. According to \cite{Will:2004}, such a streamwise tension stress
should also exist in the SCZ, corresponding to the statistical alignment of the tangled field
with the $\phi$ direction. Note that for the purposes of determining the stress, there is no
distinction between the vectors $B$ and $-B$. So, it is quite possible to have a mean stress, even an
anisotropic one such as I am considering here, without a mean field.

Formally, the streamwise stress in viscoelastic models for the turbulent magnetic stress
comes through a term proportional to the second Rivlin-Eriksen tensor. This tensor
is quite simple, all components 
being zero except for the azimuthal normal component, i.e. the $\phi\phi$ component.
In a laboratory fluid, this stress, 
and so the normal stress difference
$N_1$ as well, is equal to the product of viscous stress $\nu \gamma$ and the Weissenberg number $s \gamma$,
although strictly this depends somewhat on how one chooses to define the relaxation time $s$.
Here the situation is slightly different. Physically it makes sense to treat the turbulent magnetic field and the
turbulent velocity fluctuations on different footings, and this complicates matters. There is now a total
viscous stress which is the sum of the Reynolds viscous stress and the magnetic viscous stress, and there
is an overall effective relaxation rate which combines the Reynolds stress relaxation rate and the turbulent
Maxwell stress relaxation rate. In order to keep things simple here, let us just say that
if we are not too far away from equipartition, the effect of viscoelasticity is to introduce a
stress that is equal to $\nu s \gamma^2$ times some constant of order unity.

The curl of the divergence for such a stress is potentially nonzero and is a source for the meridional circulation.
For a stress of the form of an axisymmetric dyad $a \hat \phi \hat \phi$ such as we have here,
 the curl of the divergence is most 
simply expressed in cylindrical coordinates:
\begin{equation}
\nabla \times \left( \nabla \cdot (a \hat \phi \hat \phi) \right) = -\partial_z\left( \frac{1}{R} a \right)
\end{equation}
In the northern hemisphere in the mid-latitudes, the tendency is for
the overall magnitude of the shear to decrease with $z$. Since the elastic stress is a positive, tension stress, the
result is that the turbulent elasticity is a source for positive vorticity, which then drives a clockwise circulation.
This is the correct direction to balance the inertially-driven CCW circulation, in principle.

Let us estimate the stress needed to perform this balancing act and see if it is reasonable.
One way of describing the magnitude of the required stress is to compare it to the viscous stress. 
Using a reference value of $\nu = 10^{13.5}\,\mathrm{cm^2/s}$
and $\gamma \simeq R \Omega' \simeq R (\Delta \Omega) / (\Delta R) \simeq 10^{-6.5}\mathrm{/s}$, the viscous stress
is $\gamma \nu \simeq 10^{7}\,\mathrm{cm^2/s^2}$. The ratio of the required turbulent streamwise stress to the turbulent
viscous stress is then
\begin{equation}
\frac{\nu s \gamma^2}{\nu \gamma} = s \gamma \simeq 10^2.
\label{eq:we}
\end{equation}
A current estimate for the ratio
of streamwise stress to viscous stress for
turbulence driven by the magnetorotational instability (MRI) in accretion disks is that it is of the order of $10^{0.5}$ to $10^{1}$, based
on shearing-sheet simulations {\citep{Williams:2005}}. The interior of the SCZ is another matter, but I suggest
that it is encouraging that the estimate in eq.~({\ref{eq:we}}) differs by no more than one and a half orders of magnitude 
from the MRI results.

Alternatively, 
the strength of the turbulent, toroidal magnetic field can be expressed in terms of the Alfv\'en velocity,
$v_A^2 = B^2/(4 \pi \rho)$.
We want the centrifugal source of negative (CCW) vorticity, $R \partial_z (\Omega^2)$, to be balanced by the 
elastic Faraday tension source of positive (CW) vorticity, $-R^{-1}\partial_z(v_A^2)$. Consider a point
of mid-depth in the convection zone, somewhere in the mid-latitudes. Here
\begin{equation}
R \partial_z(\Omega^2) \simeq R \Omega \frac{\Delta \Omega}{\Delta R}
\end{equation}
and
\begin{equation}
\frac{1}{R}\partial_z(v_A^2) \simeq \frac{1}{R} \frac{v_A^2}{\Delta R}.
\end{equation}
Using the fiducial values of $\Omega = 2\pi \times 400\,\mathrm{nHz}$,  $\Delta\Omega = 2\pi \times 20\,\mathrm{nHz}$,
$R = R_\odot$, the corresponding azimuthal Alfv\'en velocity is $v_A \simeq {\mathrm{few}}\times 10^4 \,\mathrm{cm/s}$,
or of the order of $v_{\mathrm{rot}} / {\mathrm{few}}$. For a characteristic density in the middle of the convection
zone of $\rho = 0.05\, \mathrm{g/cm^3}$ this results in fields of roughly $10^{1.5}\,\mathrm{kG}$. This does not seem unreasonable.

I conclude, then, that simulations of the SCZ that include the dynamic effect of
the Faraday tension of a  toroidal field --- such as through a viscoelastic turbulence model,
 direct numerical simulation of the field, or otherwise --- may find that this
makes a significant difference in the meridional circulation and angular momentum balance;
in particular, I hypothesize that the presence of a magnetic field may make the calculated angular velocity profile more
nearly coincide with the observed profile, and be an important ingredient in the solution of
the Taylor number puzzle.

\acknowledgements 
I am very grateful to the conference organizers for providing travel funding, 
which made my presentation possible.



\end{document}